\documentclass[11pt,a4paper,pdftex]{article}
\pdfoutput=1
\usepackage{jcappub}
\usepackage{color}

\newcommand{\be}{\begin{eqnarray}}
\newcommand{\ee}{\end{eqnarray}}

\title{Iron K$\alpha$ line of boson stars}

\author[a]{Zheng~Cao,}
\author[b,c]{Alejandro~C\'ardenas-Avenda\~no,}
\author[a]{Menglei~Zhou,}
\author[a,d,1]{Cosimo~Bambi,%
\note{Corresponding author}}
\author[e]{Carlos~A~R~Herdeiro,}
\author[e]{Eugen~Radu}

\affiliation[a]{Center for Field Theory and Particle Physics and Department of Physics,\\
Fudan University, 220 Handan Road, 200433 Shanghai, China}
\affiliation[b]{Programa de Matem\'atica, Fundaci\'on Universitaria Konrad Lorenz,\\ 
Carrera 9 Bis No. 62-43, 110231 Bogot\'a, Colombia}
\affiliation[c]{eXtreme Gravity Institute, Department of Physics,\\ 
Montana State University, 59717 Bozeman MT, USA}
\affiliation[d]{Theoretical Astrophysics, Eberhard-Karls Universit\"at T\"ubingen,\\ 
Auf der Morgenstelle 10, 72076 T\"ubingen, Germany}
\affiliation[e]{Departamento de F\'isica da Universidade de Aveiro and\\ 
Center for Research and Development in Mathematics and Applications (CIDMA),\\ 
Campus de Santiago, 3810-183 Aveiro, Portugal}

\emailAdd{zcao13@fudan.edu.cn}
\emailAdd{alejandro.cardenasa@konradlorenz.edu.co}
\emailAdd{mlzhou13@fudan.edu.cn}
\emailAdd{bambi@fudan.edu.cn}
\emailAdd{herdeiro@ua.pt}
\emailAdd{eugen.radu@ua.pt}

\abstract{The present paper is a sequel to our previous work [Y. Ni et al., JCAP 1607, 049 (2016)] in which we studied the iron K$\alpha$ line expected in the reflection spectrum of Kerr black holes with scalar hair. These metrics are solutions of Einstein's gravity minimally coupled to a massive, complex scalar field. They form a continuous bridge between a subset of Kerr black holes and a family of rotating boson stars depending on one extra parameter, the dimensionless scalar hair parameter $q$, ranging from 0 (Kerr black holes) to 1 (boson stars). Here we study the limiting case $q=1$, corresponding to rotating  boson stars. For comparison, spherical boson stars are also considered. We simulate observations with XIS/Suzaku. Using the fact that current observations are well fit by the Kerr solution and thus requiring that acceptable alternative compact objects must be compatible with a Kerr fit, we find that some boson star solutions are relatively easy to rule out as potential candidates to explain astrophysical black holes, while other solutions, which are neither too dilute nor too compact are more elusive and we argue that they cannot be distinguished from Kerr black holes by the analysis of the iron line with current X-ray facilities.}


\begin{document}

\maketitle

%%%%%%%%%%%%%%%%%%%%%%%%%%%%%%%

\section{Introduction}

The current paradigm for the nature of astrophysical black hole (BH) candidates is that they are well described by the Kerr metric~\cite{kerr}. Such a solution describes a rotating BH and is completely specified by only two parameters: the mass $M$ and the spin angular momentum $J$ of the compact object. On the theoretical side, this view is supported by the uniqueness theorems (see~\cite{Chrusciel:2012jk} for a review), stating that, in 4-dimensional general relativity, the only stationary, axisymmetric, asymptotically-flat, regular ($i.e.$ without geometric singularities or closed time-like curves on or outside the event horizon) solution of the vacuum Einstein equations is the Kerr metric.  Dynamically, any initial deviations from the Kerr geometry occurring during gravitational collapse, should be radiated away through the emission of gravitational waves~\cite{price}. Moreover, the only other traditionally accepted degree of freedom -- electric charge -- can be, most likely, ignored, because a neutral equilibrium  should  be reached fairly quickly due to the highly ionized host environments of these objects, rendering any residual electric charge  too small (for macroscopic bodies) to appreciably affect the spacetime metric~\cite{ec,neutralize,neutralize2}. Also, the possible presence of an accretion disk should have a negligible geometric backreaction, because the disk mass is typically many orders of magnitude smaller than the mass of the BH candidate~\cite{disk1,disk2}. In the end, deviations from the Kerr metric seem only to be possible in the presence of new physics, thus making their detection potentially very rewarding.

Consequently, in the past 10~years, there has been a significant work to study how present and future facilities can test the Kerr BH hypothesis, both with electromagnetic observations~\cite{em1,em2,em3,Berti:2015itd} and gravitational waves~\cite{Berti:2015itd,gw1,gw2,gw3}. Among the electromagnetic approaches, the analysis of the iron K$\alpha$ line in the X-ray reflection spectrum of the accretion disk is a particularly powerful and promising tool to probe the strong gravity region around BH candidates~\cite{i1,i2,i3,i4,i5,i6,i7,i8}. The shape of the iron K$\alpha$ line is significantly affected by relativistic effects (Doppler boosting, gravitational redshift, light bending) occurring in the vicinity of the compact object. In the presence of high quality data and with the correct astrophysical model, this technique promises to provide stringent constraints on possible deviations from the Kerr metric because it can break the parameter degeneracy between the spin and possible non-Kerr features~\cite{ii1,ii2,ii3}. Other techniques meet, by contrast, serious difficulties to break such a parameter degeneracy~\cite{k1,k2,k3,k4}.

Tests of the Kerr BH hypothesis fall, at least, into two general categories. The first one considers novel exact solutions of general relativity (or even modified gravity), by analysing more general matter contents beyond (electro-)vacuum, thus capable of producing non-Kerr metrics (see $e.g.$~\cite{ag2005, gp2009, m2015,hbh,Herdeiro:2015gia,Herdeiro:2015tia,Herdeiro:2016tmi}). The second one designs parametrized families of metric deformations from Kerr~\cite{p1,p2,p3,p4,p5,p6,p7,p8}, without worrying about which model they solve. Within the first approach, Kerr BHs with scalar hair (KBHsSH)~\cite{hbh} have recently gained a lot of attention due to their physically reasonable and astrophysically plausible matter sources. These solutions are regular on and outside an event horizon, and also obey all energy conditions. They are exact 
(albeit numerical)
solutions of Einstein's gravity minimally coupled to a massive, complex scalar field, and interpolate between (a subset of) Kerr BHs -- when a normalized ``hair" parameter, $q$, vanishes -- and a family of gravitating solitons~\cite{Schunck:2003kk} -- when $q$ is maximal ($q=1$) --, the so-called boson stars; see, e.g., Ref.~\cite{Liebling:2012fv} for a review. They can bypass the various no-scalar-hair theorems that apply to this model by combining rotation with a harmonic time-dependence in the scalar field --  see~\cite{Herdeiro:2015waa} for a review of such theorems. Although these solutions are found numerically~\cite{hbh}, a proof of their existence in a small neighborhood of the Kerr family is available~\cite{proof}.

In Ref.~\cite{yy}, we have computed the iron K$\alpha$ line expected in the reflection spectrum of a small sample of KBHsSH with $q<1$, to check whether present and future X-ray missions can constrain the scalar charge of astrophysical BH candidates. We found that some KBHsSH would have an iron line definitively different from those seen in the spectrum of BH candidates, and therefore such solutions are at tension with current data. Other considerably hairy BHs, however, are compatible with current data, but future X-ray mission will be able to put much stronger constraints.

This work is the continuation of the study presented in Ref.~\cite{yy}. Here we consider the limiting case with $q=1$, in which there is no horizon and the solution describes an everywhere smooth, topological trivial configuration,  $i.e.$ a boson star.
 As in Ref.~\cite{yy}, we study the iron line profile that should be expected in the reflection spectrum of the accretion disk around these objects. We simulate observations with XIS/Suzaku\footnote{http://heasarc.gsfc.nasa.gov/docs/suzaku/} to check whether it is possible to test the existence of these objects with the current X-ray facilities. We study a sample of 12~boson star solutions, whose lensing -- another property of phenomenological interest -- was considered in~\cite{Cunha:2015yba}. We find that some solutions can be surely ruled out, as Kerr models cannot provide a good fit of their iron line, while they do it with current X-ray data. Other solutions, assuming certain emissivity profiles, can have an iron line too similar to that of Kerr BHs to be tested by current X-ray missions. Last, there are a few solutions that may be tested by current X-ray facilities, but we cannot conclude they can be ruled out by current data without a more sophisticated analysis. We remark that boson stars have been suggested as BH mimickers, and studies in this context, including of the comparative phenomenology, have been report in, $e.g.$~\cite{Mielke:2000mh,Burt:2011pv,Guzman:2005bs,Berti:2006qt,Guzman:2009zz,Eilers:2013lla,Marunovic:2013eka,Vincent:2015xta,Meliani:2015zta,Meliani:2016rfe,v2-1,v2-2}.

The content of the paper is as follows. In Section~\ref{s-2}, we briefly review the boson star solutions.  We then describe the sample of 12~solutions that shall be studied in this paper. In Section~\ref{s-3}, we compute the shape of the iron line profile expected in the reflection spectrum of these 12~spacetimes. In Section~\ref{s-4}, we simulate observations with XIS/Suzaku and we study whether these boson starts solutions can be tested with present X-ray observatories. Summary and conclusions are presented in Section~\ref{s-5}. Throughout the paper, we employ natural units in which $c = G_{\rm N} =\hbar = 1$.

\section{Boson stars \label{s-2}}
John Wheeler famously put forward the idea of ``geons"~\cite{Wheeler:1955zz}, proposing that 
%all 
elementary particles admitted a fundamental description as geometric-electromagnetic entities, made purely of gravitational and electromagnetic fields. Unfortunately, within classical Einstein-Maxwell theory, no everywhere regular, stationary, asymptotic flat solutions describing localized lumps of energy exist~\cite{Heusler:1996ft}, at least with trivial topology, rendering Wheeler's vision unphysical
(such configurations, however, could be found for other asymptotics~\cite{Melvin:1963qx,Herdeiro:2015vaa}). 

Still, Wheeler's proposal motivated much further work. In particular, Kaup decided to search for ``geons" in scalar-vacuum, rather than electrovacuum~\cite{Kaup:1968zz}. In order to avoid no-go Derrick-type theorems, Kaup included a time dependence for the scalar field, which, for it to be compatible with a static geometry, was chosen as a harmonic time-dependence, as for stationary states in quantum mechanics, and the scalar field was taken to be complex. Then, for such a complex scalar field, if a mass term is present, the corresponding Einstein-Klein-Gordon model admits geon-type solutions (also constructed in~\cite{Ruffini:1969qy}), which in modern language are called \textit{boson stars}, and regarded as an example of a gravitating soliton. 
In a nutshell, the harmonic time dependence provides an effective pressure that can balance the system's self-gravity and prevent gravitational collapse into a BH, up to a maximal total mass, that depends on the scalar field mass. 

Over the years, several generalizations of the original boson stars have been discussed, considering, for instance, self-interactions for the scalar field, and many of their physical properties have been analysed -- 
see~\cite{Jetzer:1991jr,Schunck:2003kk,Liebling:2012fv} for reviews. 
In particular, a subset of boson stars are stable against perturbations and have been shown to form dynamically; rotating, stationary boson stars (rather than static, spherical ones) have also been constructed, starting with~\cite{Schunck:1996he,Yoshida:1997qf} 
(see also~\cite{Ryan:1996nk,Kleihaus:2005me,Kleihaus:2007vk,Grandclement:2014msa,Herdeiro:2015tia,Mielke:2016war,Kan:2016xkn}) 
and analogous solutions made up of massive vector (rather than massive scalar) fields have, furthermore, recently been shown to exist~\cite{Brito:2015pxa}. 

Boson star solutions in the Einstein-(massive, complex)Klein-Gordon system are found by taking, besides the appropriate (static or stationary) ansatz for the metric, a scalar field ansatz of the form:
\begin{equation}
\Phi(t,r,\theta,\varphi)=\phi(r,\theta)e^{-iwt+im\varphi} \ ,
\end{equation}
where $\phi$ is the profile function, $w$ is the scalar field harmonic frequency and $m\in \mathbb{Z}^+$ is the azimuthal harmonic index; also, $r,\theta,\varphi$ are spherical-type coordinates, whereas $t$ is the time coordinate.  Spherical boson stars have $m=0$ and the scalar profile only depending on the radial coordinate, $r$. For $m\geqslant 1$ we have rotating boson stars. Here we shall focus only on the $m=1$ case. 
Also, all solutions to be considered here are nodeless, meaning that the scalar field profile has no nodes. 

\begin{figure}[h!]
\begin{center}
\includegraphics[type=pdf,ext=.pdf,read=.pdf,width=12cm]{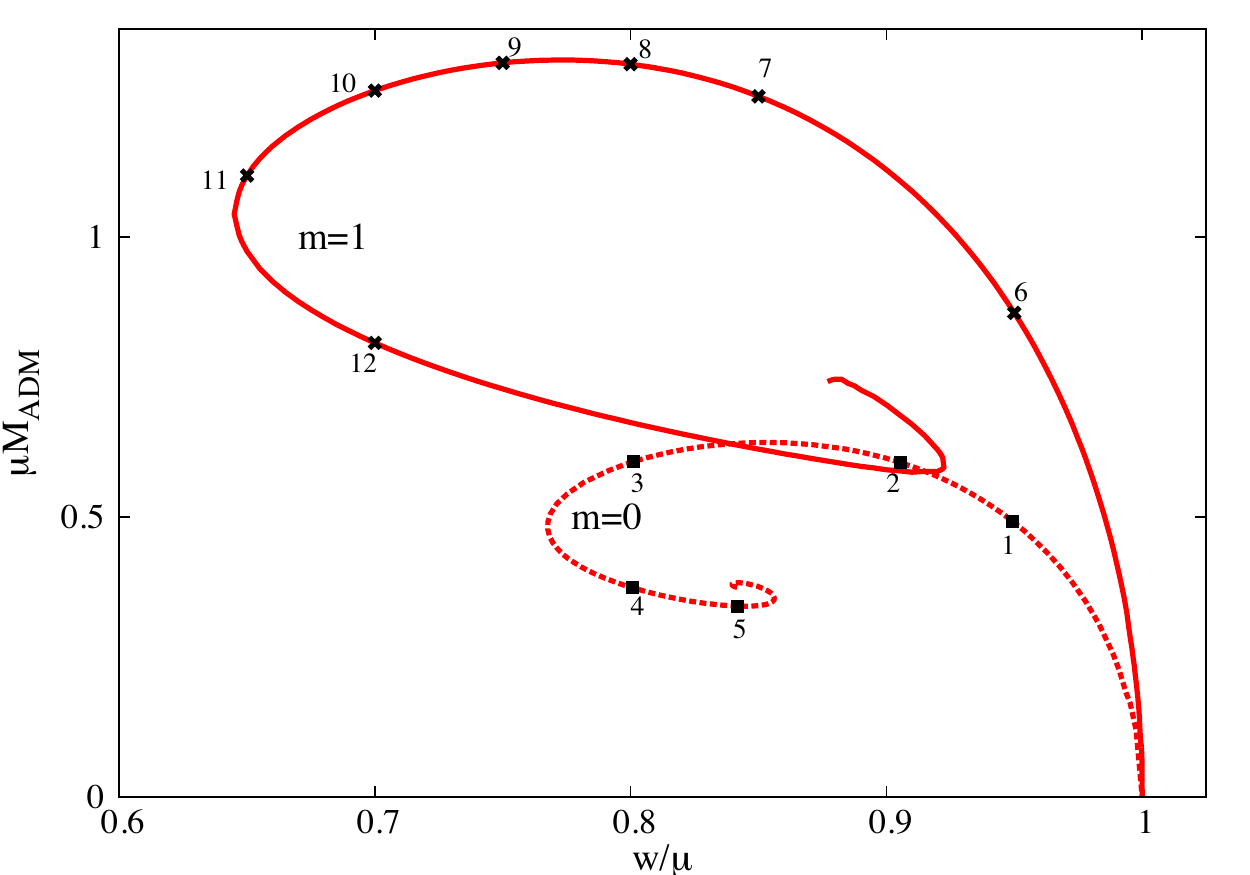}
\end{center}
\vspace{-0.5cm}
\caption{Domain of existence of the boson star solutions in an ADM mass $vs.$ scalar field frequency diagram. The red solid (dashed) line describes the family of rotating boson stars with $m=1$ (spherical boson stars with $m=0$). The twelve highlighted points correspond to the configurations to be analysed below. The same spiraling pattern holds for  higher $m$, $cf.$ Fig. 1 (left) in~\cite{Herdeiro:2015gia}.\label{domain}}
\end{figure}

An informative way to see the domain of existence of these boson stars is an ADM mass $M$ $vs.$ scalar field frequency diagram, as that shown in Fig.~\ref{domain}.  
In this plot, the red solid curve describes the rotating  boson star solutions whereas the red dashed curve describes the static ones. Overall, 12 particular solution points are highlighted corresponding to the specific solutions to be analized below, whose basic properties are described in Tab.~\ref{tab1}.  These are the same solutions that have been analysed in the context of lensing, in~\cite{Cunha:2015yba}. The axes are shown in (Planck units and) units of the scalar field mass $\mu$. Thus, the vertical axis may be regarded as (half of) the Schwarzschild radius of the boson star over the Compton wave length of the scalar field.  One can see that it cannot be much larger than unit. In other words, when there is too much scalar mass, such that the  Compton wave length of the scalar field is much smaller than the corresponding Schwarzschild radii, collapse cannot be avoided, and there are no boson star solutions. 

When following the $(w,M)$ spiral, one can distinguish different regions. First, there is a region of weak gravity,  
with $w$ close
to $\mu$  (solutions 1 and 6),
where the solutions are quite spread out, with an effective size much larger than their Schwarzschild radii, and a Newtonian description is accurate. Moving further along either of the spirals, the compactness of the boson stars increases. After the maximal ADM mass is attained, the solutions are expected to become unstable against perturbations~(see $e.g.$~\cite{Herdeiro:2015gia}). Thus, solutions 3,4,5 and 9 to 12 are expected do be unstable, even though, in the rotating case, the time scales are not precisely known. Also, when the solutions are sufficiently compact, they develop light rings. For the rotating case this happens for solutions 10-12. In the spherical case, this only occurs beyond solution 5 and hence none of the spherical solutions possess light rings.  
We also remark that the solutions can be uniquely labeled by the maximal value, $\phi_{max}$, of the scalar profile $\phi$, which increases monotonically along the boson star curve (with $\phi_{max} \to 0$ as $w\to \mu$)~\cite{Herdeiro:2015gia}.

\begin{table}
\begin{center}
\begin{tabular}{|c|cccc|c|ccc|}
\hline
Solution & $\hspace{0.05cm}$ $m$ $\hspace{0.05cm}$ & $w$ & $M$ & $J$ & $r_{\rm ISCO}$ & $r_1$ & $r_2$ & $r_3$ \\
\hline
\hline
1 & 0 & $\hspace{0.05cm}$ 0.95 $\hspace{0.05cm}$ & 0.490 & 0 & 0 & -- & -- & -- \\
2 & 0 & 0.90 & 0.605 & 0 & 0 & -- & -- & -- \\
3 & 0 & 0.80 & 0.597 & 0 & 0 & -- & -- & -- \\
4 & 0 & 0.80 & 0.375 & 0 & 0 & -- & -- & -- \\
5 & 0 & 0.84 & 0.341 & 0 & 0 & -- & -- & -- \\
\hline
\hline
6 & 1 & 0.95 & $\hspace{0.05cm}$ 0.864 $\hspace{0.05cm}$ & 0.878 & 4.970 & -- & -- & -- \\
\hline
7 & 1 & 0.85 & 1.250 & 1.303 & 2.152 & 8.778 & 10.585 & -- \\
8 & 1 & 0.80 & 1.308 & 1.372 & 1.574 & 5.035 & 10.779 & -- \\
9 & 1 & 0.75 & 1.310 & 1.375 & 1.150 & 3.183 & 10.289 & -- \\
\hline
10 & 1 & 0.70 & 1.261 & 1.306 & $\hspace{0.05cm}$ 0.803 $\hspace{0.05cm}$ & 1.470 & 3.640 & 9.680 \\
11 & 1 & 0.65 & 1.109 & 1.080 & 0.432 & 0.628 & 3.230 & 8.534 \\
12 & 1 & 0.70 & 0.811& $\hspace{0.05cm}$ 0.627 $\hspace{0.05cm}$ & 0.134 & 0.165 & 2.440 & 6.380 \\
\hline
\end{tabular}
\end{center}
\caption{Properties of the boson star solutions~1-12. The sixth columns (in units of $\mu$) report the value of the ISCO radius: the non-rotating solutions~1-5 have no ISCO (circular orbits are always stable), while the rotating solutions~6-12 have an ISCO, which is the same for corotating and counterrotating orbits. However, in the case of counterrotating orbits, the solutions~7-9 have stable orbits for $r$ between $r_{\rm ISCO}$ and $r_1$, as well as for $r > r_2$, while the orbits are unstable for $r$ between $r_1$ and $r_2$. In the case of the solutions~10-12, counterrotating orbits are stable for $r$ between $r_{\rm ISCO}$ and $r_1$, there are no circular orbits (stable or unstable) for $r$ between $r_1$ and $r_2$, there are unstable circular orbits for $r$ between $r_2$ and $r_3$, 
and circular orbits are again stable for $r > r_3$. 
\label{tab1}}
\end{table}

\section{Iron K$\alpha$ line \label{s-3}}

Broad iron K$\alpha$ lines are a common feature in the X-ray spectrum of astrophysical BH candidates~\cite{r1,r2,r3,r4}. In its rest-frame, this is a very narrow line around 6.4~keV for neutral iron, and shifts up to 6.97~keV for H-like iron ions. The line in the X-ray spectrum of BH candidates is instead broad and skewed due to special and general relativistic effects (Doppler boosting, gravitational redshift, light bending) occurring in the strong gravity region near the compact object. It is thus thought that, if properly understood, the iron K$\alpha$ line can be a powerful probe to test the metric of the strong gravity region~\cite{ii1,ii2,ii3}.

Within the disk-corona model~\cite{corona1,corona2}, the compact object is surrounded by a thin accretion disk, which is described by the Novikov-Thorne model~\cite{nt}. The disk is in the equatorial plane, perpendicular to the spin of the central body. The particles of the gas follow nearly geodesic circular orbits. The inner edge of the disk is at the ISCO radius $r_{\rm ISCO}$. When the particles of the gas reach the ISCO, they quickly plunge onto the central object without emitting additional radiation. The disk emits as a black body locally, and as a multi-color black body when integrated radially. The temperature of the inner edge of the disk is $\sim 1$~keV in the case of stellar-mass BH candidates and $\sim 10$~eV for supermassive BH candidates. The ``corona'' is a hotter ($\sim 100$~keV), usually optically-thin, electron cloud, which enshrouds the disk. Due to inverse Compton scattering of the thermal photons from the disk off the hot electrons in the corona, the latter becomes an X-ray source with a power-law spectrum. Some X-ray photons of the corona illuminate the disk and produce a reflection component with some emission lines. The iron K$\alpha$ line is usually the most prominent feature in the reflection spectrum of the disk.

In the Novikov-Thorne model, the particles of the gas follow nearly geodesic equatorial circular orbits, and therefore the angular velocity of the disk at any radius is given by the corresponding Keplerian angular velocity $\Omega_{\rm K}$. In the Kerr spacetime and in many other spacetimes, $d\Omega_{\rm K}/dr < 0$, and the magneto-rotational instability (MRI), which is the standard mechanism to drive the accretion process in Keplerian disks, can work. However, there are examples of non-Kerr spacetimes where there are at least regions with stable equatorial circular orbits where $d\Omega_{\rm K}/dr < 0$ does not hold~\cite{v2-4,v2-5,v2-6}. In such a case, the presence of a Keplerian disk at those radii would only be possible in the presence of an alternative mechanism to generate the necessary viscosity effects to enable the accretion of matter.

In our boson star spacetimes, it is not straightforward to check the sign of $d\Omega_{\rm K}/dr$, because we have the metric in numerical form. For corotating disks, the condition $d\Omega_{\rm K}/dr < 0$ for $r > r_{\rm ISCO}$ usually holds. Among the 12~solutions discussed here, only for the solution~6 the condition is not always satisfied for radii $r > r_{\rm ISCO}$ (but it is satisfied for sufficiently large radii). For the solution~6, the existence of a Keplerian thin disk with the inner edge at the ISCO radius requires the presence of some other mechanism to drive the accretion process at very small radii. In what follows, when we consider the configuration~6, we will assume that such a mechanism exists.

The shape of the iron K$\alpha$ line in the X-ray reflection spectrum of the disk is determined by the metric around the compact object, the inclination angle of the disk with respect to the line of sight of the observer $i$, the geometry of the emission region, and the emissivity profile of the disk. The emission region is the accretion disk, from its inner edge $r_{\rm in}$ to some outer edge $r_{\rm out}$.

The emissivity profile plays an important role in the final iron line shape and could be calculated theoretically if we knew the exact geometry of the corona. 
However, this is not the case at the moment, and therefore it is common to employ phenomenological descriptions. The simplest possibility is to model the intensity profile with a power law, i.e. to assume that the specific intensity of the radiation in the rest-frame of the gas in the accretion disk is $I_{\rm e} \propto 1/r^\alpha$, where $r$ is the radius of the emission point and $\alpha$ is the emission index. The latter can be a free parameter to be determined by the fit. A more sophisticated model is a broken power-law, i.e.
\be\label{eq-broken}
I_{\rm e} \propto \left\{
\begin{array}{ll}
1/r^{\alpha_1} \, , \quad & \text{for } r < r_{\rm b} \, , \\
1/r^{\alpha_2} \, , \quad & \text{for } r > r_{\rm b} \, .
\end{array} \right.
\ee
Here we may have three free parameters: two emissivity indexes, $\alpha_1$ and $\alpha_2$, and the broken radius $r_{\rm b}$. It is also common to assume two free parameters, $\alpha_1$ and $r_{\rm b}$, and set $\alpha_2 = 3$, which corresponds to the Newtonian limit at large radii in the lamppost geometry.

\begin{figure}
\begin{center}
\includegraphics[type=pdf,ext=.pdf,read=.pdf,width=8cm]{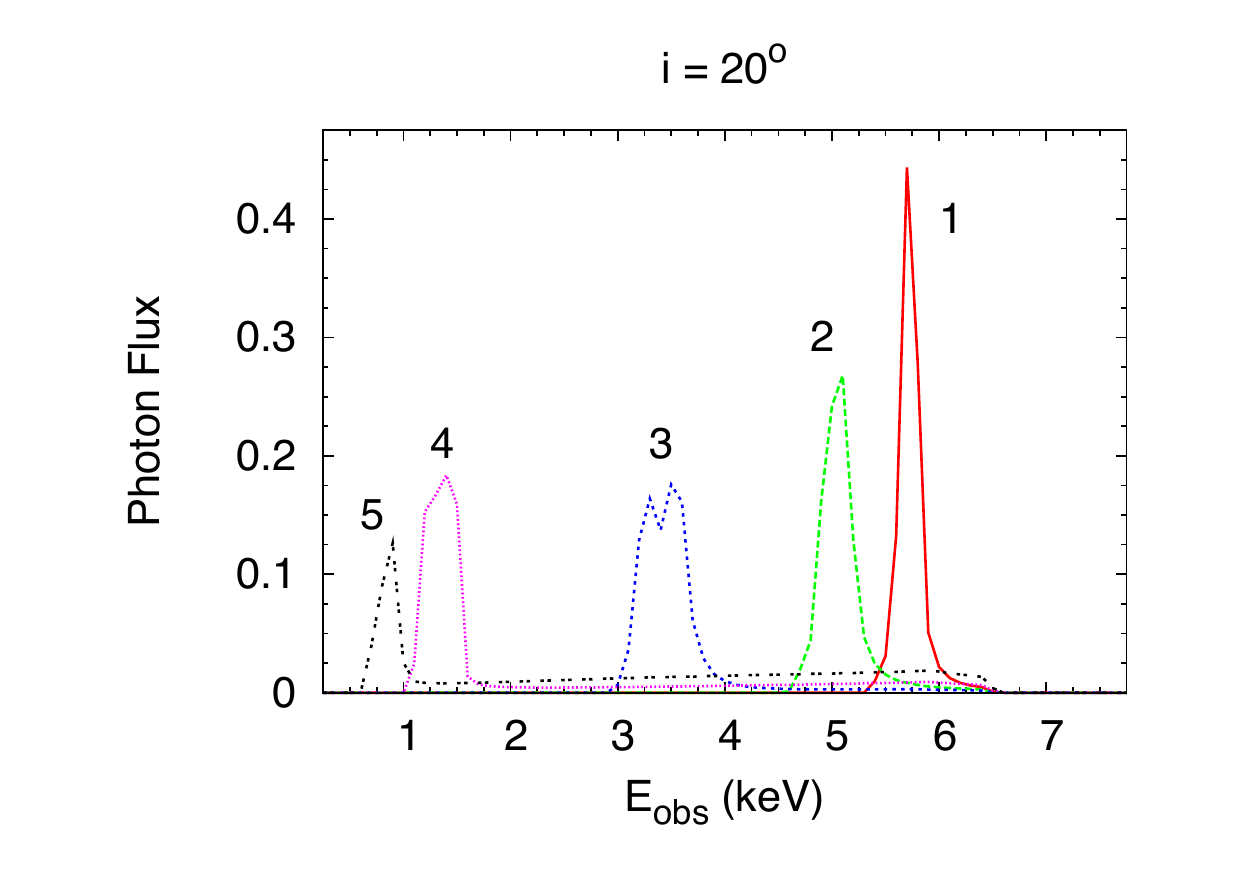} \hspace{-1.0cm}
\includegraphics[type=pdf,ext=.pdf,read=.pdf,width=8cm]{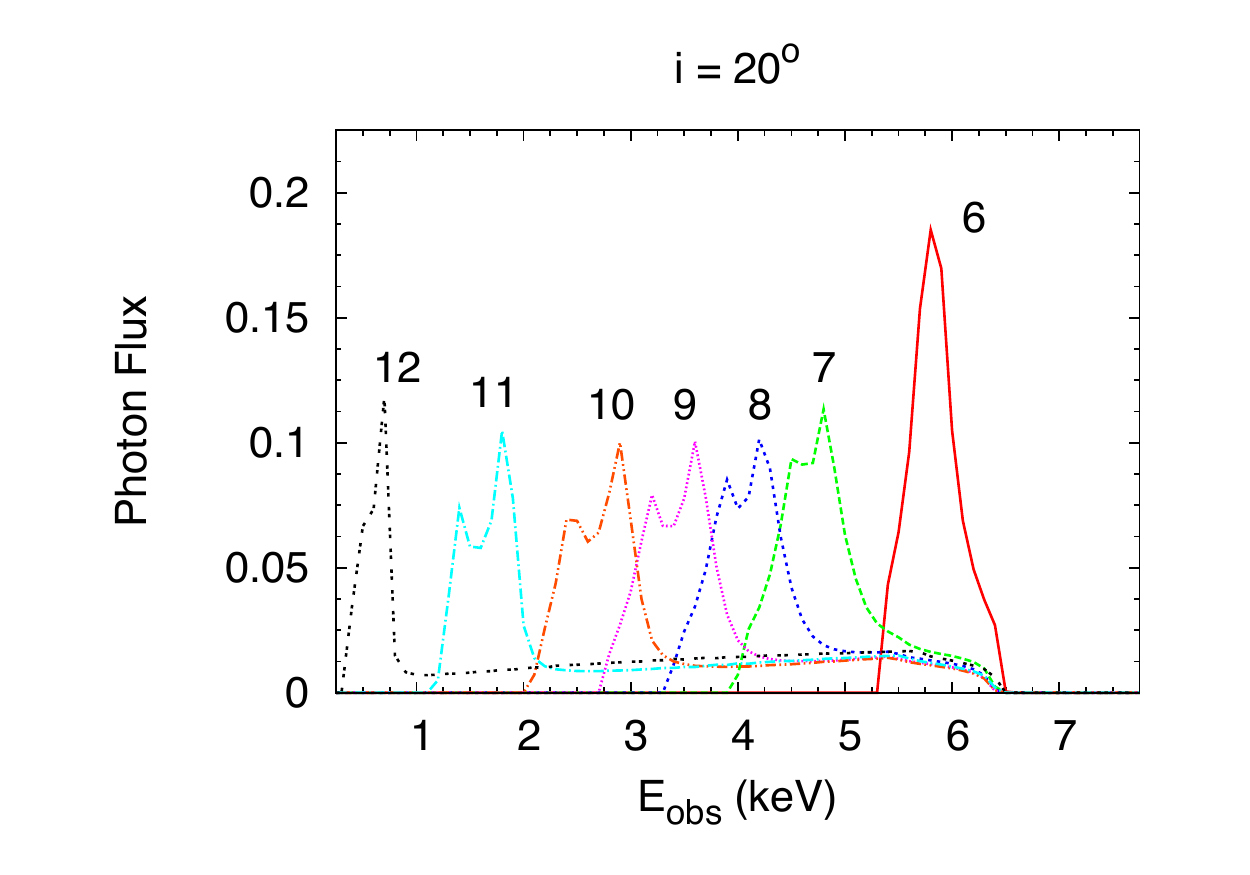} \\
\includegraphics[type=pdf,ext=.pdf,read=.pdf,width=8cm]{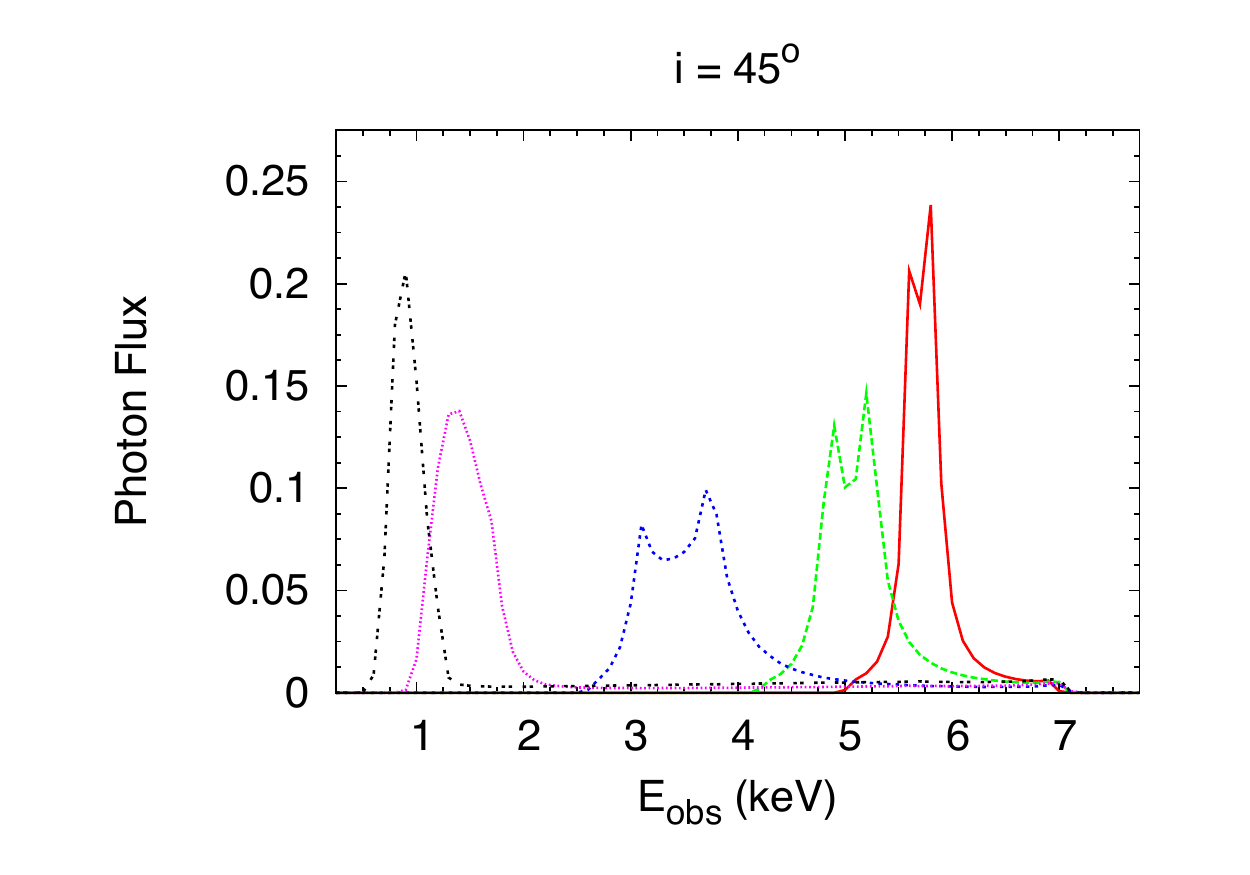} \hspace{-1.0cm}
\includegraphics[type=pdf,ext=.pdf,read=.pdf,width=8cm]{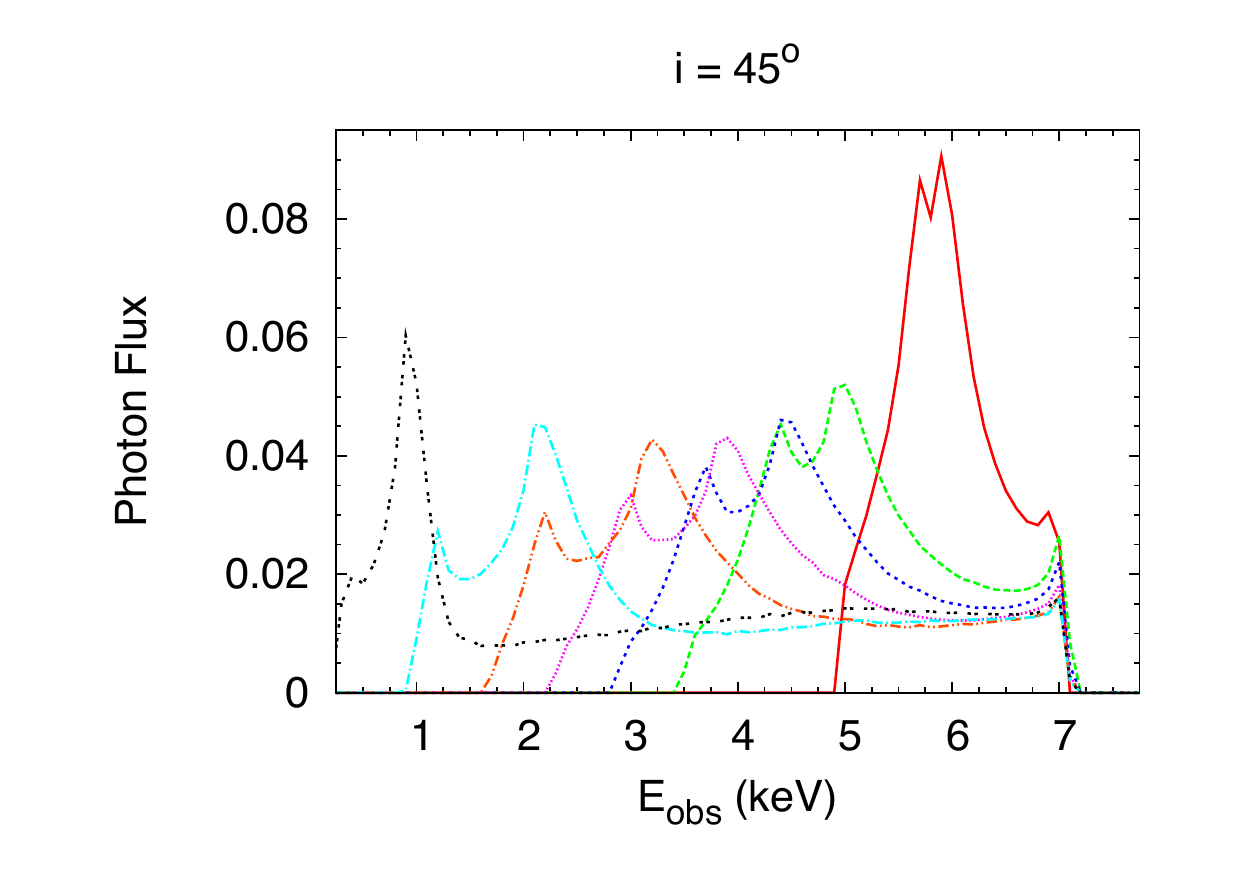} \\
\includegraphics[type=pdf,ext=.pdf,read=.pdf,width=8cm]{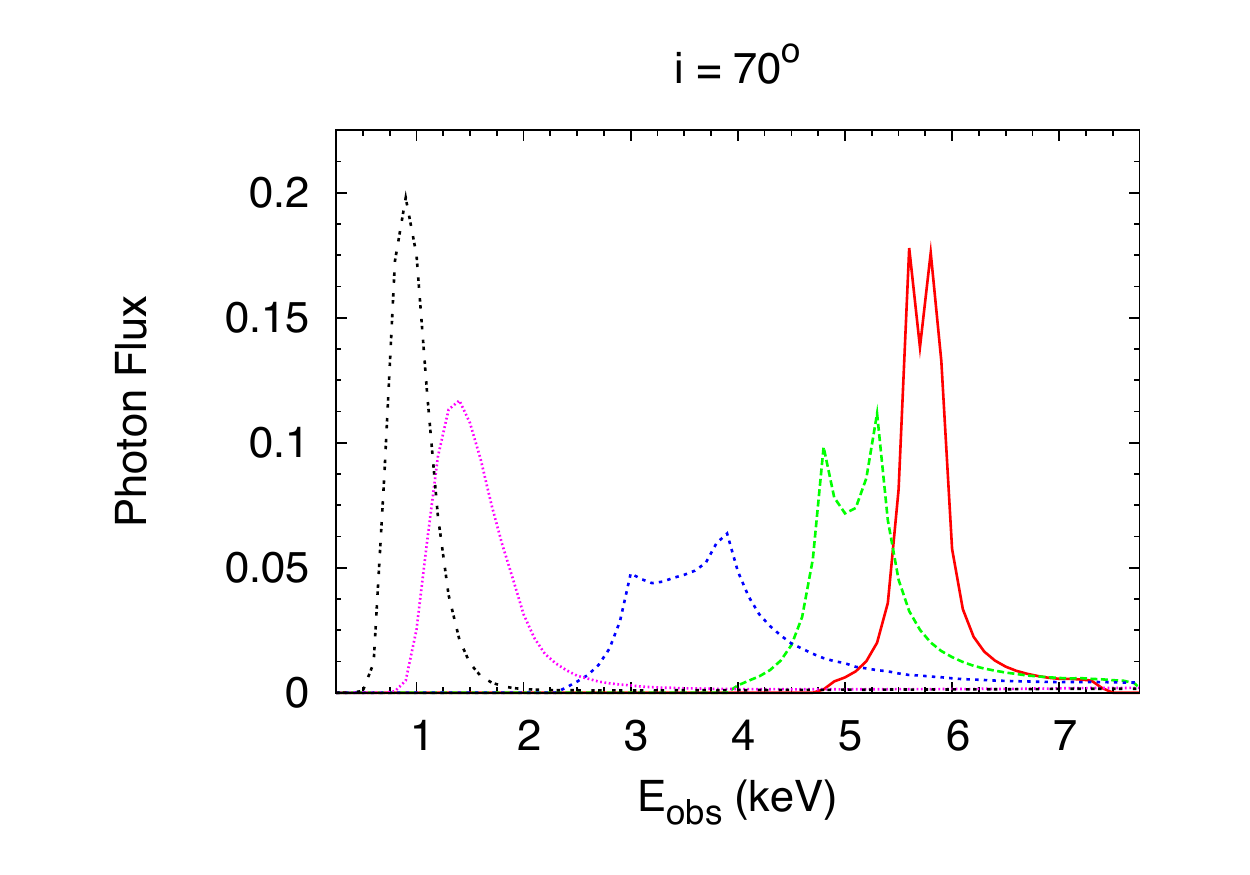} \hspace{-1.0cm}
\includegraphics[type=pdf,ext=.pdf,read=.pdf,width=8cm]{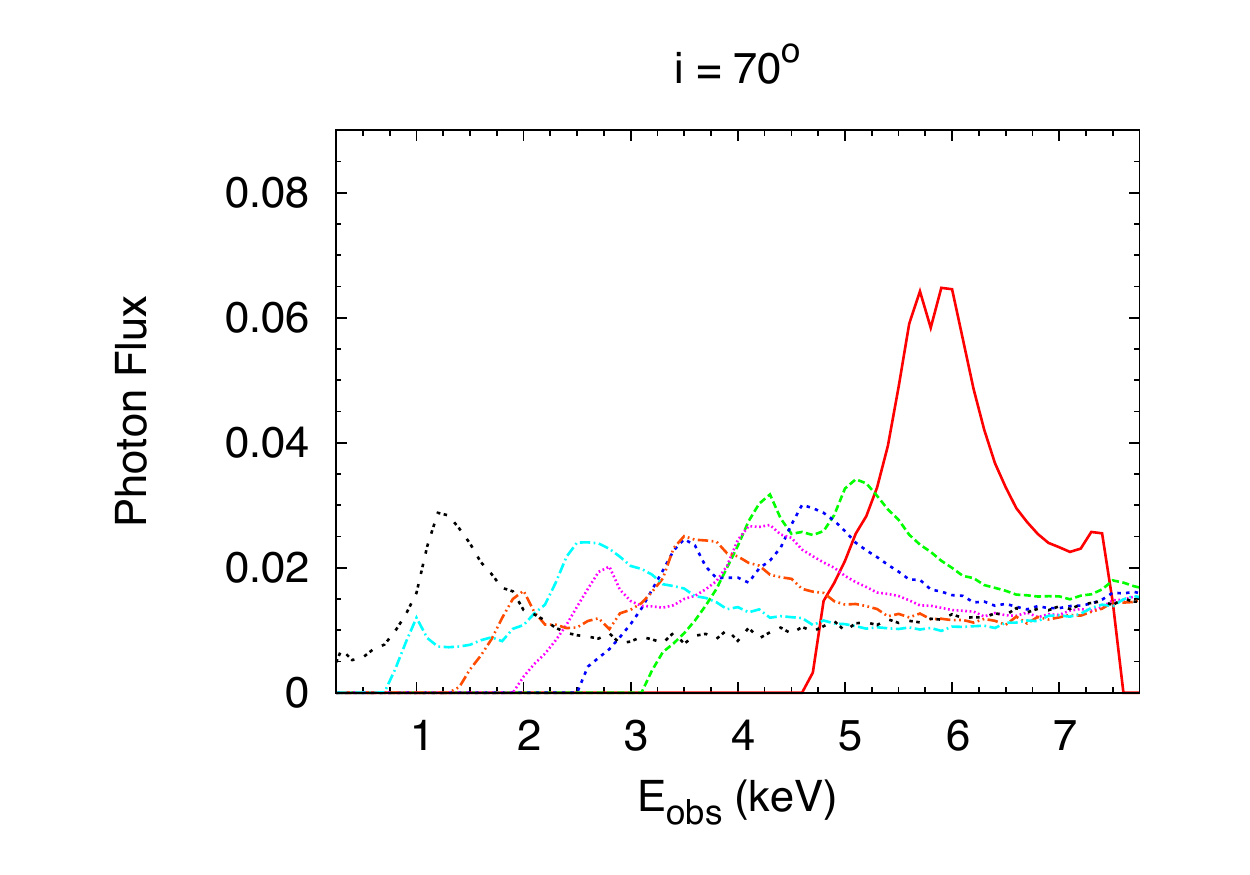}
\end{center}
\vspace{-0.5cm}
\caption{Broad iron K$\alpha$ lines expected in the X-ray reflection spectrum of the accretion disks around the boson star solutions~1-5 (left panels) and 6-12 (right panels). The inclination angle of the disk with respect to the line of sight of the observer is $i = 20^\circ$ (top panels), $45^\circ$ (central panels), and $70^\circ$ (bottom panels). The local spectrum has the intensity profile $1/r^3$. The numbers in the top panels indicate the boson star solution. For every solution, we use the same line style in the central and bottom panels. Photon flux in arbitrary units. 
\label{f-lines}}
\end{figure}

We calculate the iron lines expected in the reflection spectrum of the accretion disks of the boson star solutions 1-12 with the code described in Refs.~\cite{code,i4} and extended in Ref.~\cite{menglei} for numerical metrics\footnote{Our calculations do not include possible ghost images of the accretion disk. If the central object is a BH, the effect, if any, is very small. If the central object is not a BH, the effect might be more important~\cite{v2-3}. In the semi-quantitative analysis in the next section, we can only distinguish iron lines significantly different from those expected in the Kerr metric, and therefore we do not expect that the contribution from ghost images can alter our results and conclusions.}. The results are shown in Fig.~\ref{f-lines}. The viewing angle is $i=20^\circ$ (top panels), $45^\circ$ (central panels), and $70^\circ$ (bottom panels). The local intensity of the radiation is $I_{\rm e} \propto 1/r^3$. For the non-rotating solutions 1-5 without ISCO, we have assumed $r_{\rm in} = 0$. For the rotating solutions 6-12, we have $r_{\rm in} = r_{\rm ISCO}$. Here and in what follows, we always assume that the disk is correlating with the spin of the central object. In all these simulations, we have $r_{\rm out} = 25$ in units in which $1/\mu = 1$. In order to have $r_{\rm out}$ in terms of the gravitational radius of the object ($r_{\rm g} = M$), it is necessary to divide $r_{\rm out}$ by the ADM mass in the fourth column of Tab.~\ref{tab1}. Such a low value of $r_{\rm out}$ will be employed even later in the simulations with XIS/Suzaku because it is computationally quite expensive to compute the iron line profile for a larger outer radius. However, we stress that our results do not depend on the choice of $r_{\rm out}$ because the statistics in our simulations is not good enough to permit the measurement of the outer edge of the accretion disk.

\begin{figure}
\begin{center}
\includegraphics[type=pdf,ext=.pdf,read=.pdf,width=8.1cm]{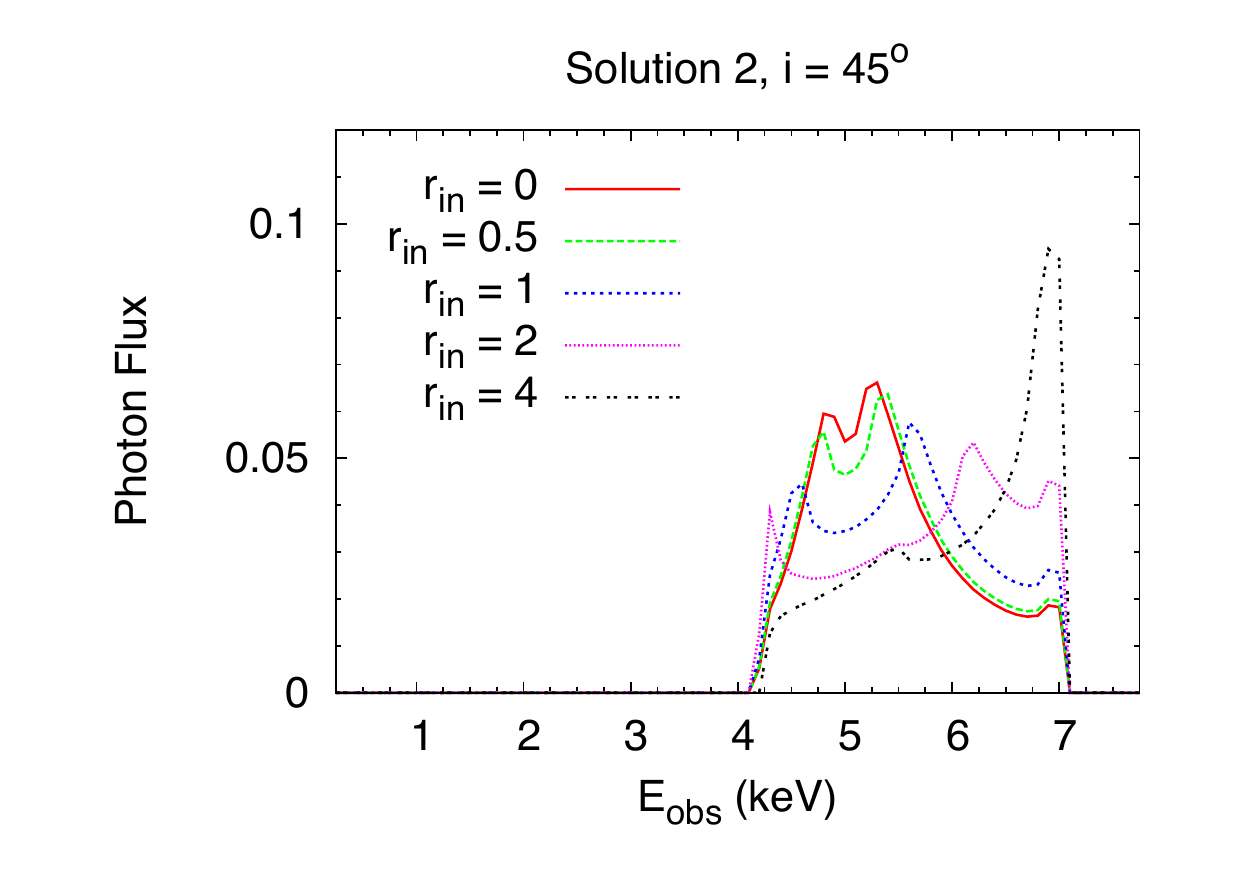} \hspace{-1.0cm}
\includegraphics[type=pdf,ext=.pdf,read=.pdf,width=8.1cm]{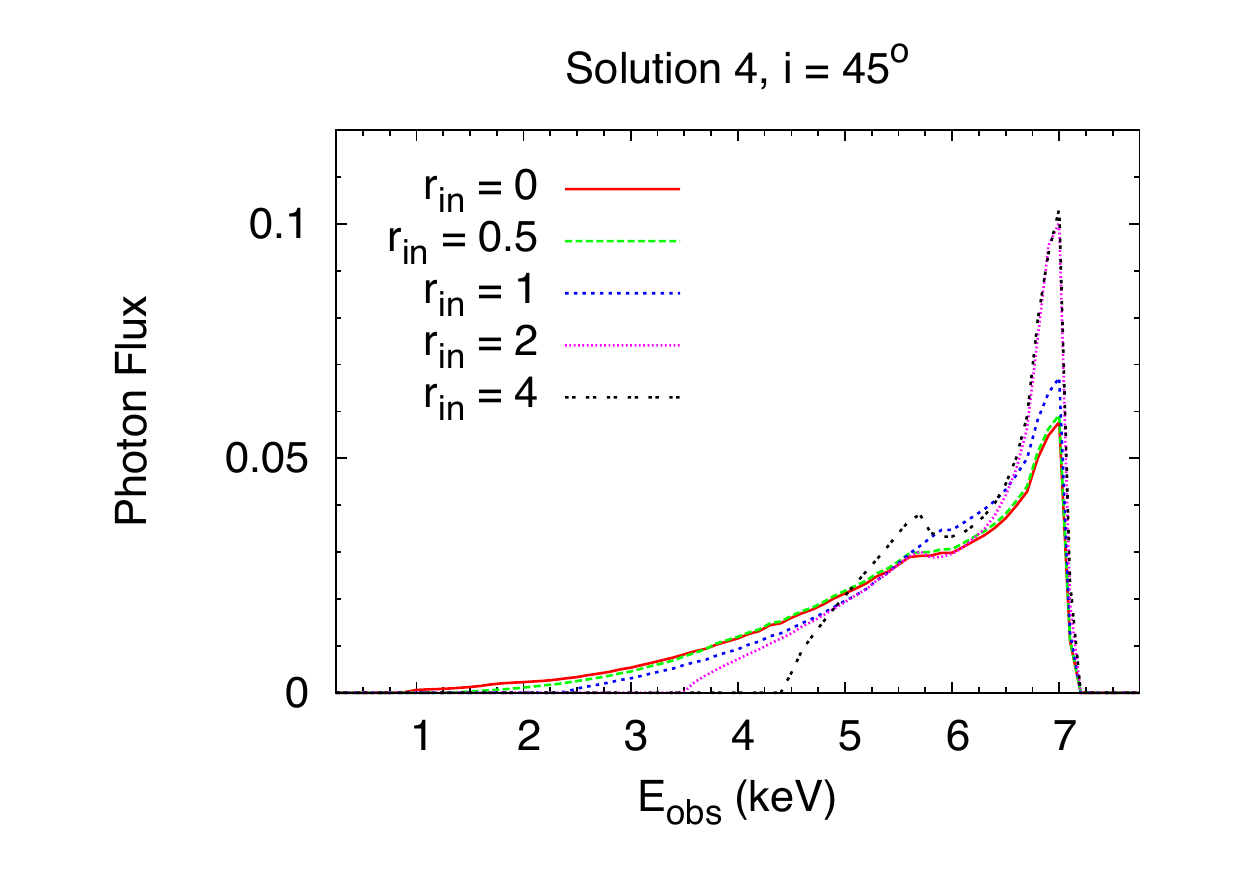}
\end{center}
\vspace{-0.5cm}
\caption{Broad iron K$\alpha$ lines expected in the X-ray reflection spectrum of the accretion disks around the boson star solutions~2 (left panel) and 4 (right panel), assuming the lamppost intensity profile $h/(h^2 + r^2)^{3/2}$ and different values of the inner radius $r_{\rm in}$. The inclination angle of the disk with respect to the line of sight of the observer is $i = 45^\circ$. Photon flux in arbitrary units. 
\label{f-lines2}}
\end{figure}

As the power-law emissivity profile $1/r^3$ is not very physical if $r_{\rm in} = 0$ or very small, we have calculated also the iron line profiles of the solutions 1-5 assuming the lamppost intensity profile in flat spacetime. In this case we have~\cite{dauser}
\be\label{eq-lamppost}
I_{\rm e} \propto \frac{h}{(h^2 + r^2)^{3/2}} \, .
\ee
Fig.~\ref{f-lines2} shows the iron line profile for the solutions~2 and 4, varying also the inner edge of the accretion disk $r_{\rm in}$ (we consider a disk truncated as in Refs.~\cite{i6,4246}). The solutions~1 and 3 are quite similar to the solution~2, while the solution~5 is similar to the solution~4. In the solutions 1-3, the photons emitted at very small radii are not strongly redshifted, so the impact on the iron line profile of a finite $r_{\rm in}$ is moderate. In the solutions~4 and 5, photons emitted at very small radii are strongly redshifted: if we change the emissivity profile from $\propto 1/r^3$ to Eq.~(\ref{eq-lamppost}), we remove the peak at low energies and the shape of the iron line looks much more like that expected in the Kerr metric.

\section{Simulations and discussion \label{s-4}}

In this section we want to understand whether the analysis of the iron line profile can distinguish boson stars and Kerr BHs, and in particular if the available X-ray data can already rule out some boson star solutions. To do this, we proceed as in Ref.~\cite{yy}. Our starting point is that the available reflection models on the market assume the Kerr metric and can provide good fits of the reflection spectrum of BH candidates observed with current X-ray facilities. We thus perform simulations of the X-ray spectrum expected in the case of the boson star solutions 1-12. We fit the simulated observations with a Kerr model. If the latter provides a good fit, we conclude that a similar observation cannot distinguish a boson star from a Kerr BH. If the fit is bad, we argue that astrophysical BHs cannot be such a boson star, because there is no tension between current observational data and theoretical reflection models in the Kerr metric. This is a simple analysis, but we expect that it is able to provide the right insight. Of course, if any future X-ray observation yields data incompatible with the Kerr model, our conclusions must be revisited.

We do not consider a specific source, but we assume some typical parameters for a bright BH binary, as a binary is a more suitable source than an AGN for this kind of tests. The X-ray spectrum of the BH binary is approximated by a single iron K$\alpha$ line and a power-law component. In principle, we should consider the whole reflection spectrum rather than a single iron line, but this is enough for a preliminary analysis, as the information on the spacetime geometry is mainly encoded in the iron line. In our simulations, the flux in the 0.7-10~keV range is about $4 \cdot 10^{-9}$~erg/s/cm$^2$ and the iron line equivalent width is about 200~eV. The photon index of the power-law component is $\Gamma = 2$. These are the same values as those employed in Ref.~\cite{yy}.

We use XSPEC\footnote{http://heasarc.gsfc.nasa.gov/docs/xanadu/xspec/index.html} with the background, the ancillary, and the response matrix files of XIS/Suzaku to simulate the data. For every boson star solution, we simulate four sets of observations: the time exposure can be either 100~ks or 1~Ms, while the emissivity of the iron line is calculated either assuming the power-law profile $\propto 1/r^3$ or the lamppost profile in Eq.~(\ref{eq-lamppost}). The photon count in the 0.7-10~keV range turns out to be about $3 \cdot 10^7$ and $3 \cdot 10^8$, respectively for the 100~ks and 1~Ms observations. For every set of observations, we consider three possible viewing angles, $i = 20^\circ$, $45^\circ$, and $70^\circ$. However, we will see that the viewing angle cannot qualitatively change the fit and thus plays only a marginal rule in our discussion.

The simulated data are fitted using XSPEC with a power-law plus a single iron line. The latter is introduced with KERRCONV*GAUSSIAN, which assumes the Kerr metric and the broken power-law profile in Eq.~(\ref{eq-broken}). For the power-law, we have two free parameters: the photon index of the power-law $\Gamma$ and the normalization of the power-law. For the iron line, we have six free parameters: the BH spin $a_*$, the inclination angle of the disk with respect to the line of sight of the observer $i$, the emissivity index $\alpha_1$, the emissivity index $\alpha_2$, the breaking radius $r_{\rm b}$, and the normalization of the iron line.

In Ref.~\cite{yy} we also simulated observations with LAD, which is expected on board of the future X-ray mission eXTP~\cite{extp}. Here we do not consider LAD/eXTP, because its statistics is so good that we would need the correct (or at least a plausible) emissivity profile for the iron line. However, in the case of the boson stars without ISCO we do not have a good model for the intensity profile and we thus prefer to omit the discussion of possible tests with LAD/eXTP. We also note that the outer radius of the accretion disk in our simulations is set to 25 (see previous section), while we fit the simulated data with an iron line model assuming a fixed outer radius $r_{\rm out} = 400$~$M$. We have checked that such a simplification does not appreciably change our fits for the simulated observations of 100~ks and 1~Ms with XIS/Suzaku. On the contrary, the exact value of the outer radius is important with the high quality data of LAD/eXTP. The effective area at 6~keV of XIS/Suzaku is about 800~cm$^2$, to be compared with the $\sim 35,000$~cm$^2$ of LAD/eXTP, and this makes the difference.

\subsection{Non-rotating boson stars $(m=0)$}

\begin{figure}
\begin{center}
\includegraphics[type=pdf,ext=.pdf,read=.pdf,width=12.5cm]{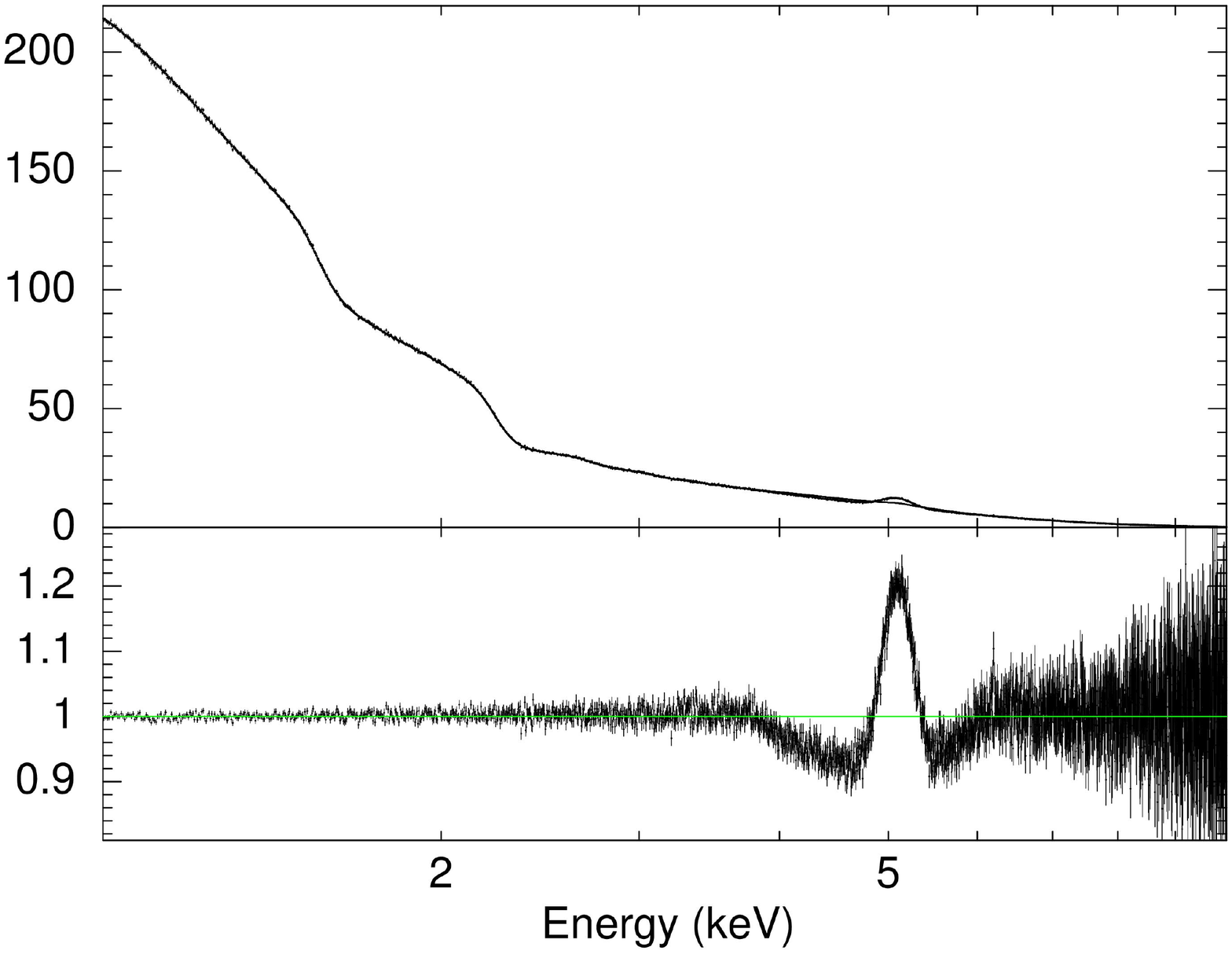}
\end{center}
\vspace{-0.5cm}
\caption{Folded spectrum and ratio between the folded spectrum and the simulated data for the boson star solution~2, assuming the viewing angle $i = 20^\circ$ and the emissivity profile $1/r^3$. The exposure time is 100~ks and the minimum of the reduced $\chi^2$ is about 6. See the text for more details. \label{f-s2i20}}
\vspace{-0.8cm}
\begin{center}
\includegraphics[type=pdf,ext=.pdf,read=.pdf,width=12.5cm]{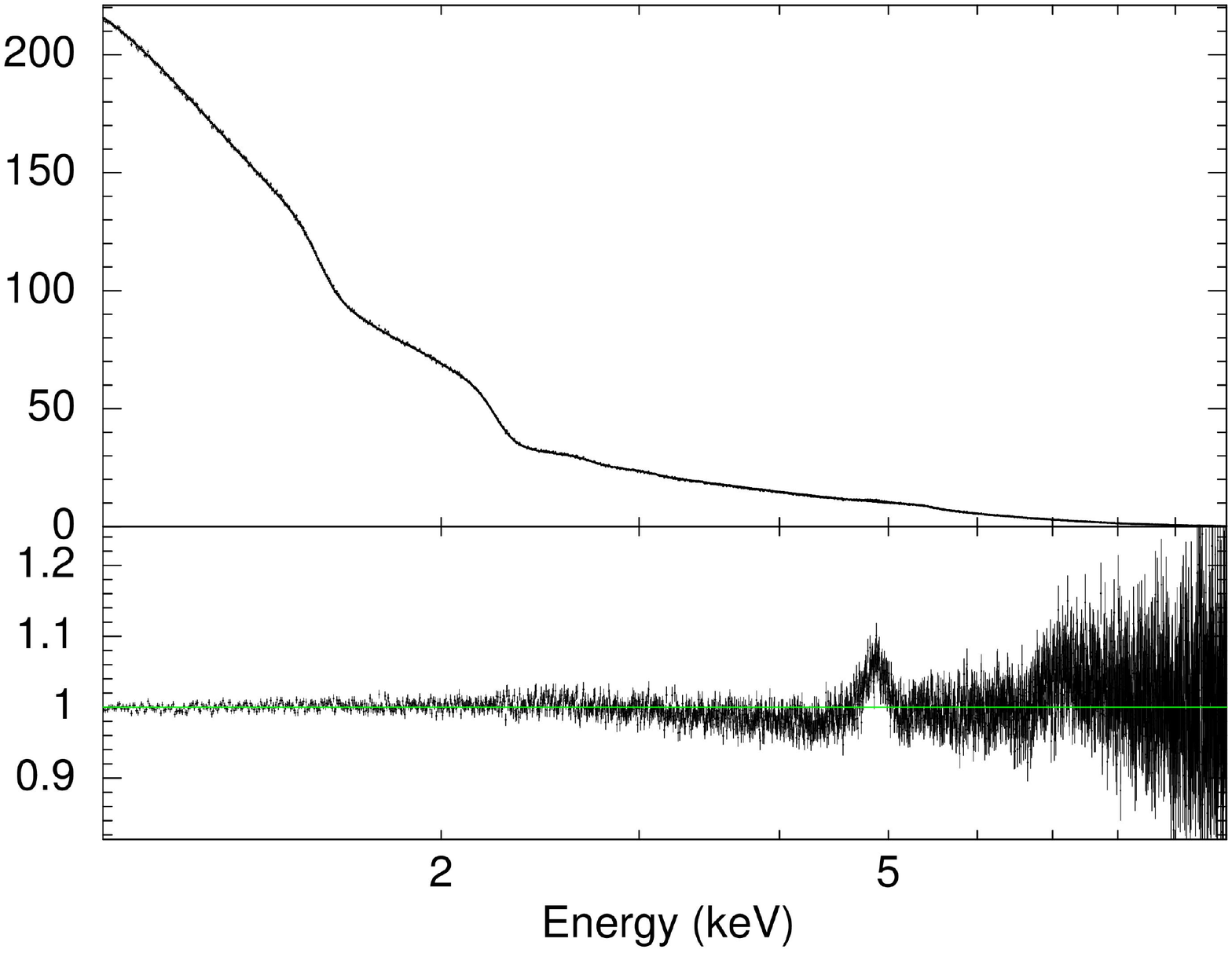}
\end{center}
\vspace{-0.5cm}
\caption{As in Fig.~\ref{f-s2i20}, but for the viewing angle $i = 70^\circ$. The minimum of the reduced $\chi^2$ is about 2 and we still observe unresolved features. See the text for more details. \label{f-s2i70}}
\end{figure}

We start simulating observations for the non-rotating boson star solutions 1-5. In the first set of simulations, the exposure time is 100~ks and the emissivity profile is modeled by the power-law $1/r^3$. These solutions have no ISCO, namely circular orbits are always stable. We first consider the case in which there is no inner edge of the disk, in the sense that we set $r_{\rm in} = 0$ in the code. We find that the Kerr model always provide a very bad fit and it seems that the fit gets worse moving along the $q = 1$ curves from the solution 1 to the solution 5. For a given solution, we find that the minimum of the reduced $\chi^2$ is lower for $i = 70^\circ$ and higher for $i = 20^\circ$. Figs.~\ref{f-s2i20} and \ref{f-s2i70} show the case of the solution~2, respectively for $i = 20^\circ$ and $i = 70^\circ$. In each figure, the top panel is the folded spectrum, the bottom panel is the ratio between the folded spectrum and the simulated data. For both viewing angles, there is a clear unresolved feature around 5~keV, suggesting that we are using the wrong model to fit the data. The fits of the solutions 1-3 are bad, with the minimum of the reduced $\chi^2$ ranging from 2 to 10. The fits of the solutions~4 and 5 are even worse, and we find that the minimum of the reduced $\chi^2$ is more than 40.

If we set the inner edge at a finite radius, like $r_{\rm in} = 0.5$ or 1, the picture changes. The fits of the solutions~1-3 are almost unaffected and still very bad. The fits of the solutions 4 and 5 become acceptable and we do not see clear unresolved features. Such a result could have been expected from the iron line profiles computed in the previous section. If $I_{\rm e} \propto 1/r^3$ and $r_{\rm in} = 0$, the iron line profile is essentially determined by the emission near the center $r = 0$. The iron line profiles of the solutions 1-3 are not strongly redshifted, despite most of the emission comes from the region near the center. In this case, if we set the inner edge of the disk at a finite radius, the iron line profile changes, but the effect is not so dramatic. The iron line profiles of the solutions 4 and 5 are very redshifted and we have a peak at low energies, because the profile $1/r^3$ diverges at $r=0$. Setting $r_{\rm in}$ at a finite radius, we remove the peak, and the fit gets better.

We then consider simulations with longer exposure time, 1~Ms. In this case, the Kerr model can never provide an acceptable fits for the solutions 1-5, no matter we change the inner edge of the disk at some finite radius.

\begin{figure}
\begin{center}
\includegraphics[type=pdf,ext=.pdf,read=.pdf,width=12.5cm]{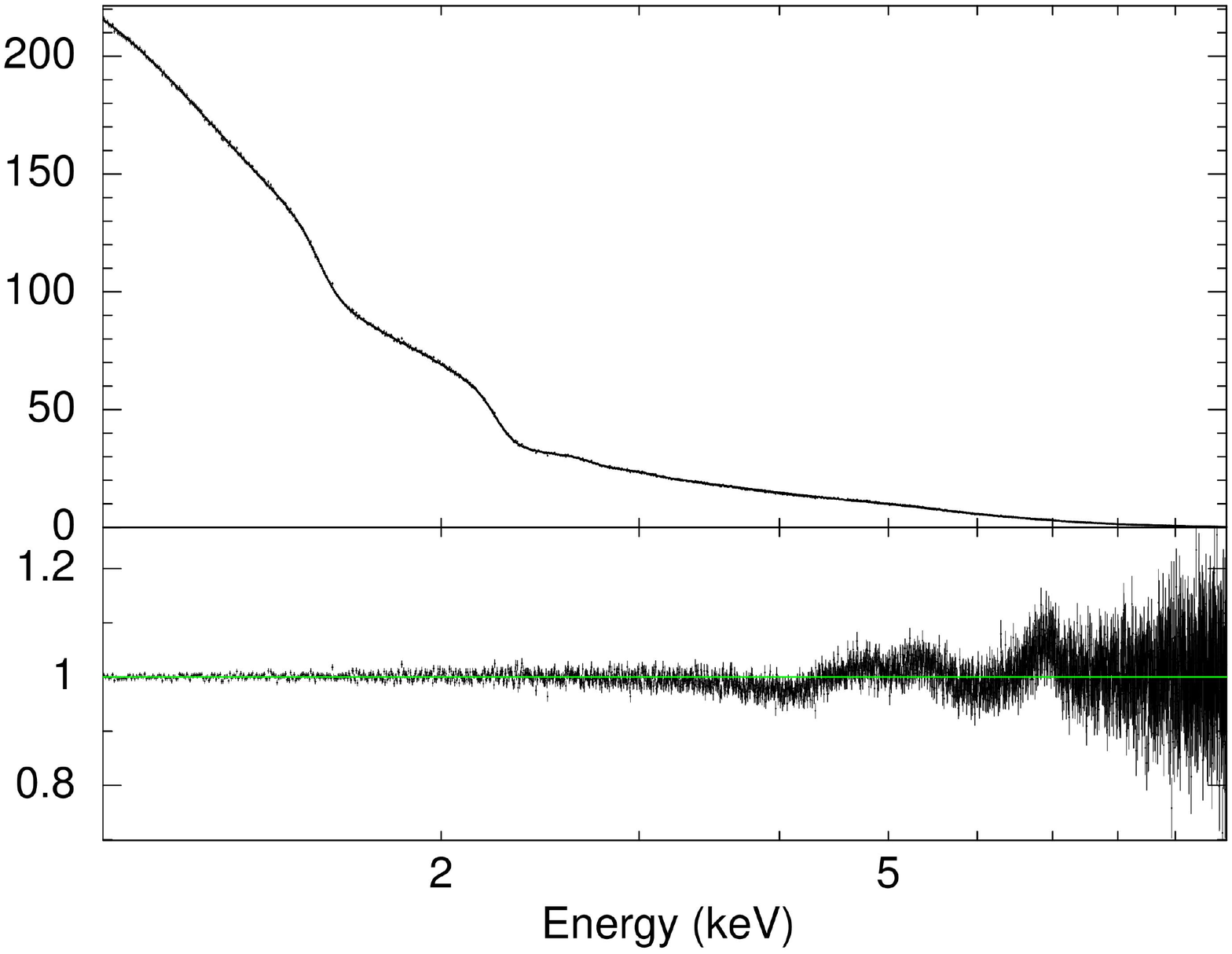}
\end{center}
\vspace{-0.5cm}
\caption{Folded spectrum and ratio between the folded spectrum and the simulated data for the boson star solution~2, assuming the viewing angle $i = 45^\circ$ and the lamppost intensity profile $h/(h^2 + r^2)^{3/2}$. The exposure time is 100~ks and the minimum of the reduced $\chi^2$ is about 1.5. See the text for more details. \label{f-s2i45t100-lamp}}
\vspace{-0.4cm}
\begin{center}
\includegraphics[type=pdf,ext=.pdf,read=.pdf,width=12.5cm]{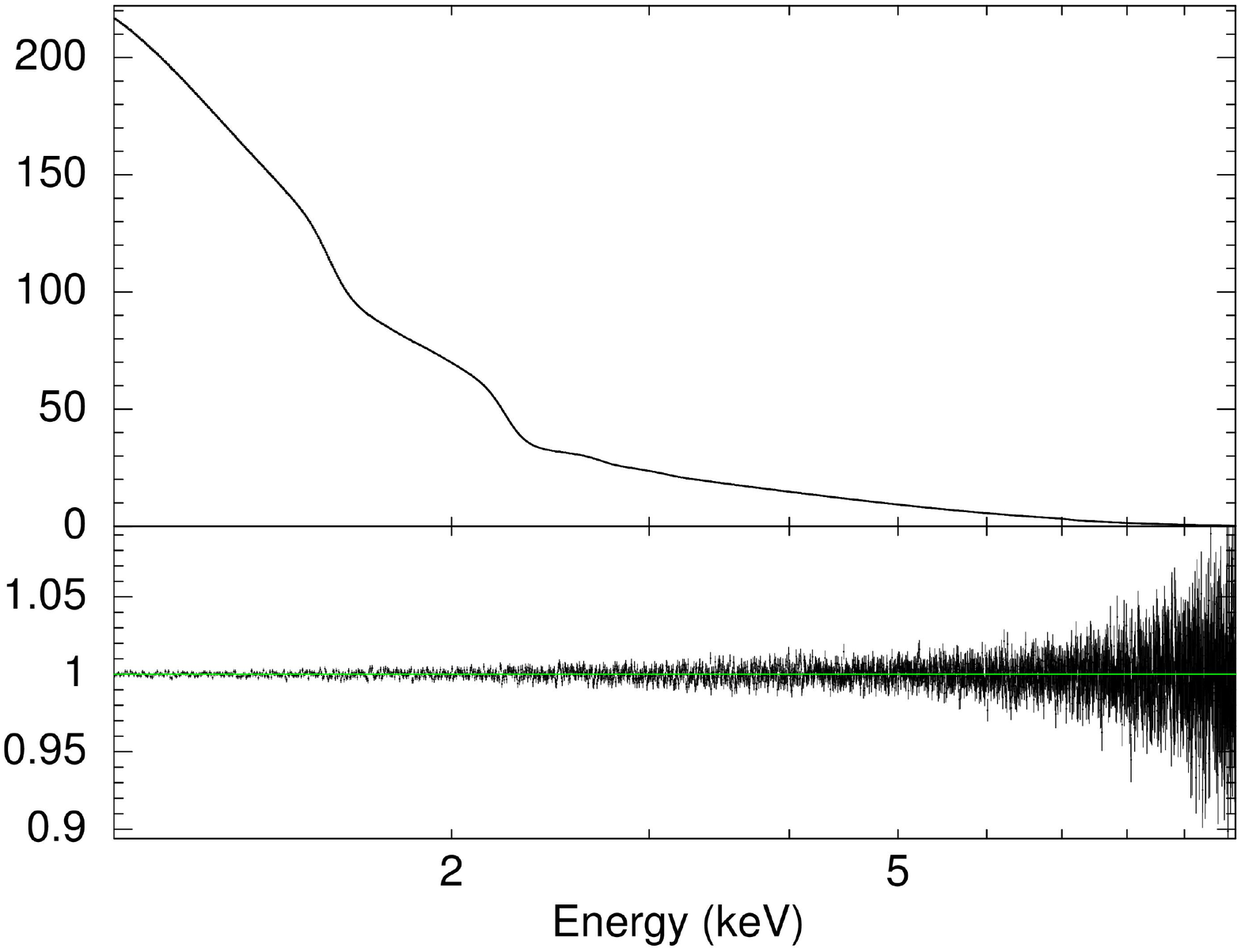}
\end{center}
\vspace{-0.5cm}
\caption{Folded spectrum and ratio between the folded spectrum and the simulated data for the boson star solution~4, assuming the viewing angle $i = 45^\circ$ and the lamppost intensity profile $h/(h^2 + r^2)^{3/2}$. The exposure time is 1~Ms and the minimum of the reduced $\chi^2$ is about 1.0. See the text for more details. \label{f-s4i45t1000-lamp}}
\end{figure}

We move to the simulations in which we adopt the lamppost emissivity profile with $I_{\rm e} \propto h/(h^2 + r^2)^{3/2}$. If the exposure time is 100~ks, the fits of the solutions 1-3 are still unacceptable with clear unresolved features, independently of the location of the inner edge of the disk. Fig.~\ref{f-s2i45t100-lamp} illustrates the case of the solution~2 as an example. The fits of the solutions 4 and 5 are acceptable and without unresolved features with $r_{\rm in} = 0$.

If we consider 1~Ms observations, the Kerr model can still provides a good fit for the spectra of the solutions 4 and 5. Fig.~\ref{f-s4i45t1000-lamp} shows the results for the solutions~4 assuming $r_{\rm in} = 0$. The minimum of the reduced $\chi^2$ is around 1.0 and the fit is indeed good. We do not see any unresolved feature in the plot of the ratio between the folded spectrum and the simulated data. The iron lines in the right panel in Fig.~\ref{f-lines2} looked indeed quite similar to the Kerr ones.

\subsection{Rotating boson stars $(m=1)$}

We repeat the analysis of the previous subsection for the seven rotating boson stars, the solutions 6-12. The inner edge of the accretion disk is set at the ISCO radius, which is always finite in this case, see Tab.~\ref{tab1}. As before, we consider the three viewing angles $i = 20^\circ$, $45^\circ$, and $70^\circ$.

\begin{figure}
\begin{center}
\includegraphics[type=pdf,ext=.pdf,read=.pdf,width=12.5cm]{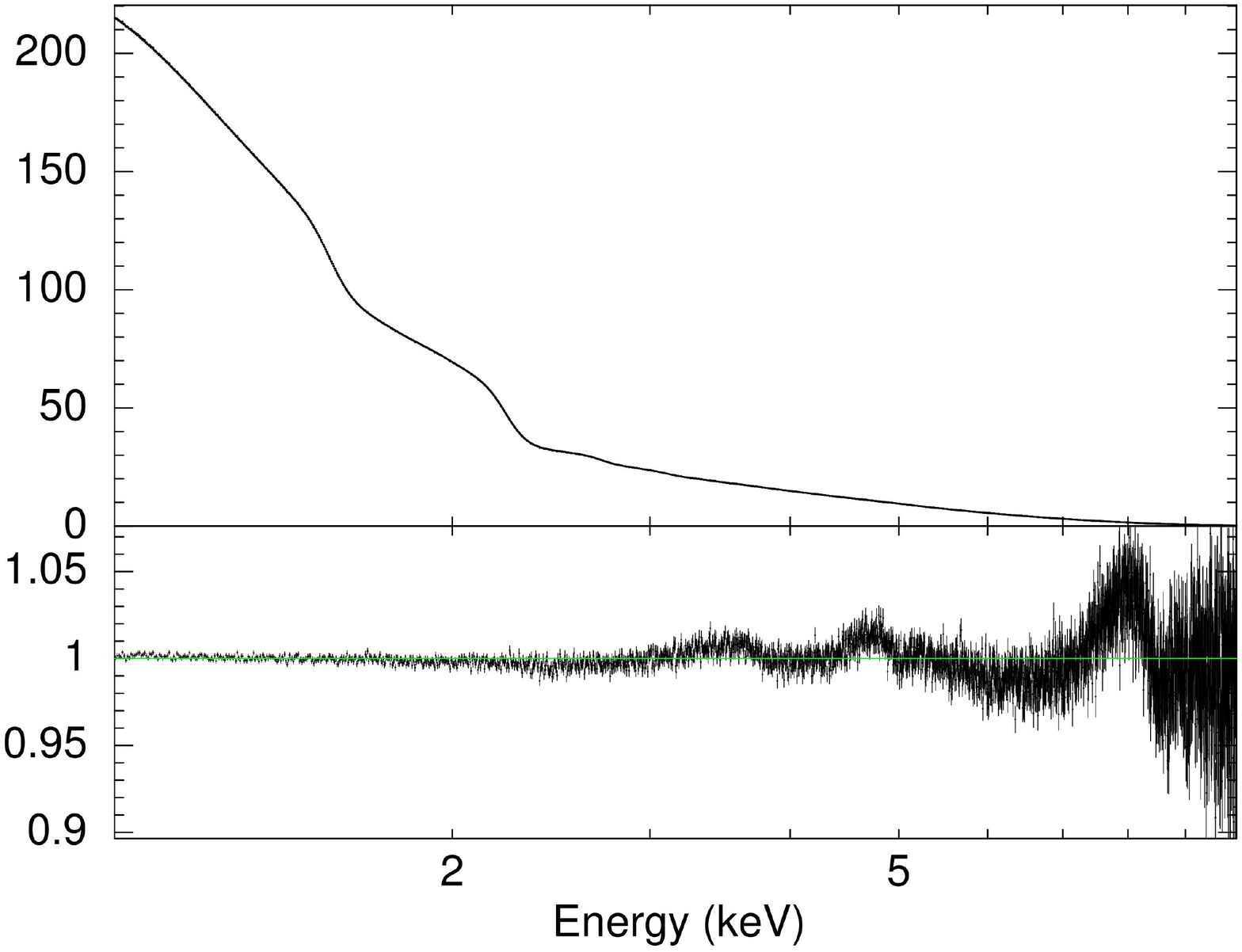}
\end{center}
\vspace{-0.5cm}
\caption{Folded spectrum and ratio between the folded spectrum and the simulated data for the boson star solution~8, assuming the viewing angle $i = 70^\circ$ and the intensity profile $1/r^3$. The exposure time is 1~Ms and the minimum of the reduced $\chi^2$ is about 2.5. See the text for more details. \label{f-s8i70t1000}}
\vspace{-0.4cm}
\begin{center}
\includegraphics[type=pdf,ext=.pdf,read=.pdf,width=12.5cm]{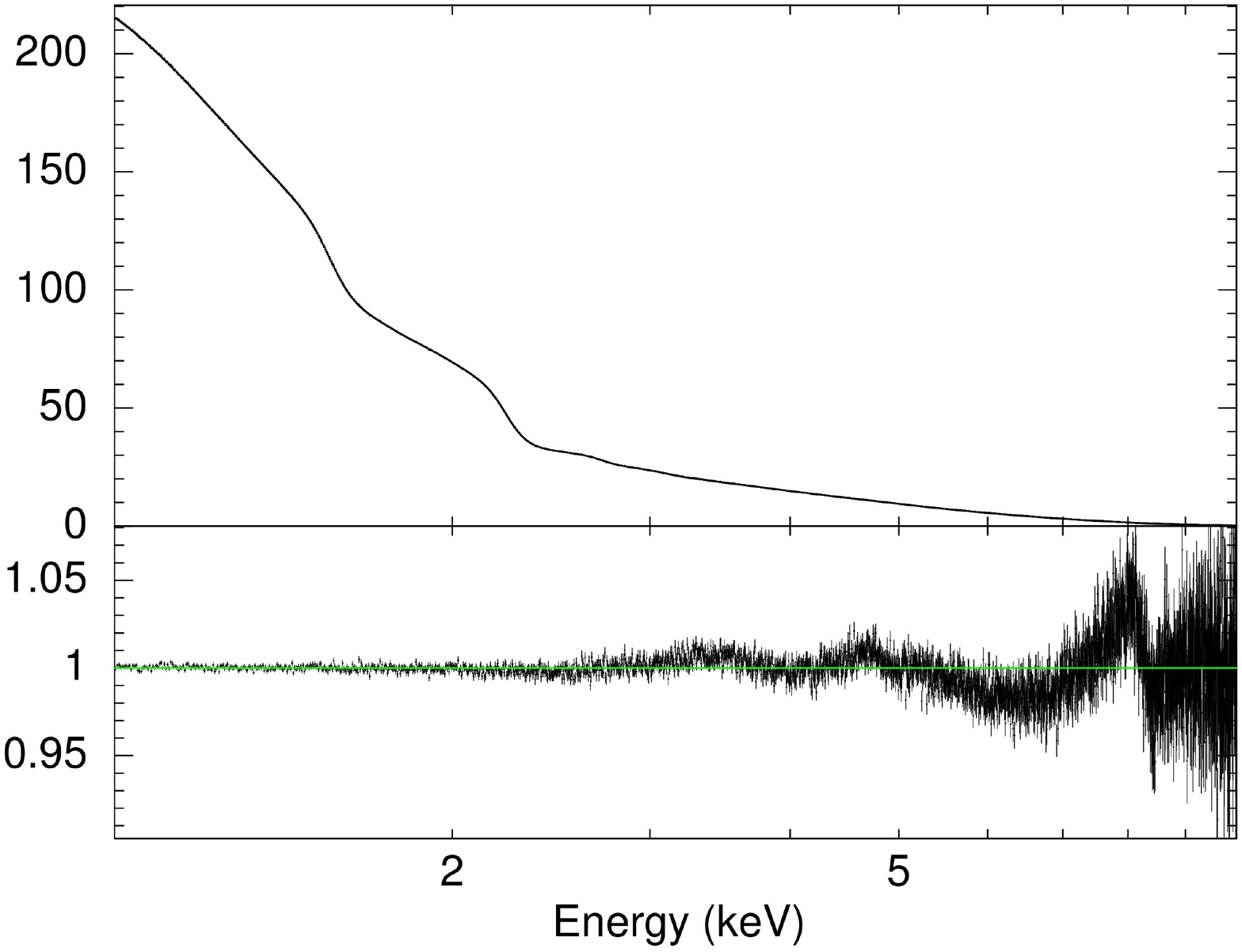}
\end{center}
\vspace{-0.5cm}
\caption{Folded spectrum and ratio between the folded spectrum and the simulated data for the boson star solution~8, assuming the viewing angle $i = 45^\circ$ and the lamppost intensity profile $h/(h^2 + r^2)^{3/2}$. The exposure time is 1~Ms and the minimum of the reduced $\chi^2$ is about 2.3. See the text for more details. \label{f-s8i45t1000-lamp}}
\end{figure}

We start assuming the power-law emissivity profile $1/r^3$ and simulate observations of 100~ks. We fit the simulated spectra with the Kerr model. The fit of the solution~8 is good: the minimum of the reduced $\chi^2$ is close to 1 and there are no unresolved features in the ratio between the folded Kerr spectrum and the simulated data. As we move along the $q=1$ curve, the fit gets worse. The fits of the solutions 7, 9, and 10 are still acceptable. The fits of the solutions 6, 11, and 12 are not. When we simulate observations of 1~Ms, we find that the Kerr model can never provide an acceptable fit to the simulated data. Fig.~\ref{f-s8i70t1000} shows the folded spectrum and the ratio between the folded spectrum and the simulated data of the solution~8. The fit is clearly bad and indeed the minimum of the reduced $\chi^2$ is around 2.5. The fits of the other solutions are worse.

We repeat the simulations for the lamppost intensity profile, in which $I_{\rm e} \propto h/(h^2 + r^2)^{3/2}$. We find the same qualitative results as the simulations with the power-law intensity profile. If the exposure time is 100~ks, the Kerr model provides a good fit for the spectrum of the solution~8, still an acceptable fit for the solutions 7, 9, and 10, while the fits is not acceptable in the case of the solutions 6, 11, and 12. Once again, when the exposure time of the simulations is 1~Ms, the fit with the Kerr model always shows some unresolved features and a large minimum of the reduced $\chi^2$. Fig.~\ref{f-s8i45t1000-lamp} shows the fit of the spectrum of the solution~8. As before, the other fits are worse.

\section{Concluding remarks \label{s-5}}

In the present paper we have continued our study to observationally test the existence in the Universe of Kerr BHs with scalar hair, which are solutions of Einstein's gravity minimally coupled to a massive, complex scalar field found in Ref.~\cite{hbh}. We have considered the limiting case $q=1$, which yields  rotating ($m=1$) boson stars. For comparison, here we have also considered non-rotating (spherically symmetric) boson stars ($m=0$).

We have computed the profile of the iron K$\alpha$ line that should be expected from the reflection spectrum of the accretion disk of these objects. We have simulated observations with XIS/Suzaku, assuming that the source is a bright BH binary, and we have fitted the spectrum with a Kerr model. Since the available X-ray data of the reflection spectrum of BH binaries are regularly fitted with Kerr models, and there is no tension between data and theoretical predictions, we argue that the BH candidates in the Universe cannot be the boson star solutions with a bad fit, while those with an acceptable fit cannot be distinguished by Kerr BHs and are thus viable candidates. We remark, again, that should any future observations of BH candidates yield spectra incompatible with the Kerr model, the conclusions herein should be revisited.

\begin{table}[h!]
\begin{center}
\begin{tabular}{|c|cc|cc|}
\hline
Solution & $1/r^3$, $t = 100$~ks & $1/r^3$, $t = 1$~Ms & LP, $t = 100$~ks & LP, $t = 1$~Ms \\
\hline
\hline
1 & $\times$ & $\times$ & $\times$ & $\times$ \\
2 & $\times$ & $\times$ & $\times$ & $\times$ \\
3 & $\times$ & $\times$ & $\times$ & $\times$ \\
4 &  & $\times$ &  &  \\
5 &  & $\times$ &  &  \\
\hline
6 & $\times$ & $\times$ & $\times$ & $\times$ \\
7 &  & $\times$ & & $\times$ \\
8 &  & $\times$ & & $\times$ \\
9 &  & $\times$ & & $\times$ \\
10 &  & $\times$ &  & $\times$ \\
11 & $\times$ & $\times$ & $\times$ & $\times$ \\
12 & $\times$ & $\times$ & $\times$ & $\times$ \\
\hline
\end{tabular}
\end{center}
\caption{Summary of our results. $\times$ means that the Kerr model fails to give an acceptable fit. In the second and the third columns, the model assumes $I_{\rm e} \propto 1/r^3$. In the fourth and the fifth columns, we have $I_{\rm e} \propto h/(h^2 + r^2)^{3/2}$. The exposure time is 100~ks (second and fourth columns) and 1~Ms (third and fifth columns). See the text for more details.  \label{tab2}}
\end{table}

Our results are summarized in Tab.~\ref{tab2}, where $\times$ is to indicate that the Kerr model fails to give an acceptable fit for the corresponding boson star solution, exposure time, and emissivity profile. The solutions 1-3, 6, 11, 12 can be ruled out. Even for an exposure time of 100~ks, we find that the Kerr model always provides a very bad fit. The iron line profiles of these boson stars are not compatible with those in the available X-ray data of BH binaries. For spherical boson stars, the qualitative conclusion is that acceptable solutions must be fairly compact. This qualitative behaviour also holds for the rotating solutions; in this case, however, both the most dilute and the most compact cases are ruled out for these ``short" exposure times. In a sense, our sample of solutions included more compact rotating than spherical boson stars, as the latter do not possess a light ring whereas the former do (beyond solution 10). Perhaps too compact spherical boson stars can also be ruled out. Unfortunately, in this case on has to go well into the spiral in order to get a light ring, which only appears in the third branch of solutions, i.e. after the second backbending in frequency, wherein the accuracy of solutions becomes poorer.

The Kerr model provides an acceptable fit for the solutions 7-10 and for an exposure time of 100~ks, but fails to give an acceptable fit in the case of a longer observation of 1~Ms. In such a case, we should investigate these solutions better and analyze real data of specific sources to conclude whether the solutions 7-10 can be excluded or are consistent with the available data. Unfortunately, this is beyond the possibilities of the current version of our code.

Last, we have the solutions 4 and 5. If we assume the lamppost emissivity profile $\sim h/(h^2 + r^2)^{3/2}$, even an observation of 1~Ms cannot distinguish these solutions from a Kerr BH, in the sense that the Kerr model provides a good fit. We thus argue that it impossible to test the existence of these objects with the current X-ray missions.

%%%%%%%%%%%%%%%%%%%%%%%%%%%%%%%

\begin{acknowledgments}

ZC, MZ, and CB were supported by the NSFC (grants 11305038 and U1531117) and the Thousand Young Talents Program. CB also acknowledges support from the Alexander von Humboldt Foundation. AC-A acknowledges funding from the Fundaci\'on Universitaria Konrad Lorenz (Project 5INV1161) and the NSF CAREER Grant PHY-1250636. CH and ER acknowledge funding from the FCT-IF programme. This work was partially supported by  the  H2020-MSCA-RISE-2015 Grant No.~StronGrHEP-690904, and by the CIDMA project UID/MAT/04106/2013.

\end{acknowledgments}

%%%%%%%%%%%%%%%%%%%%%%%%%%%%%%

\end{document}